\documentclass[
  sigconf,
  natbib=false,
  anonymous=false,
  final,
]{acmart}

\usepackage[T1]{fontenc}    %
\usepackage[utf8]{inputenc} %
\usepackage[english]{babel} %
\usepackage{microtype}      %
\usepackage[
  autostyle,
]{csquotes}                 %
\usepackage{scrhack}        %
\usepackage[
  newcommands
]{ragged2e}                 %
\PassOptionsToPackage{
  hyphens
}{url}                      %
\usepackage[
  style=ACM-Reference-Format,
  giveninits=true,          %
  backend=biber,
  backref=false,
  doi=false,
  isbn=false,
  url=false,
  eprint=false,
  sortcites=true,           %
  sorting=none,
  maxcitenames=2,
]{biblatex}                 %
\bibliography{paper_arxiv.bib}  %
\DeclareFieldFormat*{title}{%
  \mkbibemph{#1\isdot}}     %
\DeclareFieldFormat*{citetitle}{%
  \mkbibemph{#1\isdot}}     %
\usepackage{flushend}       %
\usepackage{amssymb}
\usepackage{amsmath}
\usepackage{booktabs}       %
\usepackage{tabu}
\usepackage{csquotes}
\usepackage{thmbox}
  \newtheorem[style=L,leftmargin=8pt,rightmargin=2pt]{scenario}{Scenario}
  \newtheorem{definition}{Definition}[section]
\usepackage[inline]{enumitem}
  \setlist[itemize]{nosep, noitemsep, leftmargin=1.5em}
\def\hyp{-\penalty0\hskip0pt\relax} %
\usepackage{tikz}
  \usetikzlibrary{arrows,matrix,positioning,fit,tikzmark}
  \usepackage{environ} %
\usepackage{fontawesome}
\usepackage{pgfplots}
    \pgfplotsset{compat=newest}
    \usepackage[font=footnotesize]{subfig}
\usepackage[
  colorinlistoftodos,
  textwidth=\marginparwidth,
  textsize=footnotesize,
  obeyFinal,
]{todonotes}                %
  \presetkeys{todonotes}{%
    fancyline,
    color=red!30}{}
\usepackage[                %
  capitalise,               %
  nameinlink
]{cleveref}
  \crefname{section}{Sect.}{Sect.}
  \Crefname{section}{Section}{Sections}
  \crefname{scenario}{Scenario}{Scenario}
\usepackage{nicefrac}
\makeatletter
\let\orgdescriptionlabel\descriptionlabel
\renewcommand*{\descriptionlabel}[1]{%
  \let\orglabel\label
  \let\label\@gobble
  \phantomsection
  \protected@edef\@currentlabel{#1}%
  \protected@edef\@currentlabelname{\let\hfil\relax #1}%
  \let\label\orglabel
  \orgdescriptionlabel{#1}%
}
\makeatother

\DeclareRobustCommand{\roe}{data record}
\DeclareRobustCommand{\roepl}{data records}

\DeclareRobustCommand{\reusage}{re\hyp{}{}usage}

\DeclareRobustCommand{\pseudomapping}{pseudonym lookup table}

\newcommand{\code}{\normalfont\texttt}
\newcommand{\framework}{\normalfont\textsf}

\setcopyright{acmcopyright}
\acmDOI{xx.xxx/xxx_x}

\acmISBN{978-1-4503-6866-7/20/03}

\acmConference[SAC'20]{ACM SAC Conference}{March 30-April 3, 2020}{Brno, Czech Republic} 
\acmYear{2020}
\copyrightyear{2020}

\acmArticle{4}
\acmPrice{15.00}

\renewcommand\footnotetextcopyrightpermission[1]{} %

\begin{document}

\fancyhead[RO,LE]{} %

\title{PEEPLL: Privacy-Enhanced Event Pseudonymisation with Limited
  Linkability
}
\titlenote{%
  This paper is an extended version of work that has been accepted for
  publication in the proceedings of
  the 35th ACM/SIGAPP Symposium On Applied Computing, 2020.
}

\author{Ephraim Zimmer}
\affiliation{%
  \institution{University of Hamburg}
}

\author{Christian Burkert}
\affiliation{%
  \institution{University of Hamburg}
}

\author{Tom Petersen}
\affiliation{%
  \institution{University of Hamburg}
}

\author{Hannes Federrath}
\affiliation{%
  \institution{University of Hamburg}
}
\renewcommand{\shortauthors}{E. Zimmer et al.}

\begin{abstract}
  Pseudonymisation provides the means to reduce the privacy impact of
  monitoring, auditing, intrusion detection, and data collection in general on
  individual subjects.
  Its application on \roepl{}, especially in an environment with additional
  constraints, like re-identification in the course of incident response,
  implies assumptions and privacy issues, which contradict the achievement of
  the desirable privacy level.
  Proceeding from two real-world scenarios, where personal and identifying data
  needs to be processed, we identify requirements as well as a system model for
  pseudonymisation and explicitly state the sustained privacy threats, even when
  pseudonymisation is applied. With this system and threat model, we derive
  privacy protection goals together with possible technical realisations, which
  are implemented and integrated into our event pseudonymisation framework
  \framework{PEEPLL} for the context of event processing, like monitoring and
  auditing of user, process, and network activities.
  Our framework provides privacy\hyp{}friendly linkability in order to maintain
  the possibility for automatic event correlation and evaluation, while at the
  same time reduces the privacy impact on individuals. Additionally, the
  pseudonymisation framework is evaluated in order to provide some restrained
  insights on the impact of assigned paradigms and all necessary new mechanisms
  on the performance of monitoring and auditing. 
  With this framework, privacy provided by event pseudonymisation can be
  enhanced by a more rigorous commitment to the concept of personal data
  minimisation, especially in the context of regulatory requirements like the
  European General Data Protection Regulation.
\end{abstract}


\ccsdesc[500]{Security and privacy~Pseudonymity, anonymity and untraceability}
\ccsdesc[500]{Security and privacy~Management and querying of encrypted data}
\ccsdesc[300]{Security and privacy~Public key encryption}
\ccsdesc[300]{Security and privacy~Hash functions and message authentication codes}

\keywords{personal data minimisation, pseudonym \reusage{}, privacy protection
  goals, pseudonymisation framework, indistinguishability,
  unobservability, limited linkability}

\maketitle

\section{Introduction}\label{sec:intro}

Monitoring and auditing of user, process, and network activities plays an
important role for the security of a system. By leveraging gained information, a
security operator can observe unusual behaviour and possibly detect or even
prevent past and ongoing attacks on his system. The success of such an analysis
requires as much information about the actions in the system as possible.
However, it also has severe privacy implications. Each \roe{} might not only
contain directly or indirectly identifying attributes of a person, like names or
addresses, but also IP addresses, unique user ids, or account data. Such data is
called \textit{personal data}~\cite[Article~4(1)]{GDPR16} or \textit{quasi
identifier} (QIDs)~\cite{BWJ08} and comprises directly identifying as well as
potentially identifying features, i.e., attributes that are by themselves not
sufficient to identify individuals but may in combination be used to do so.
Analysing such data constitutes a severe impact on the privacy of individuals.
It facilitates the creation of user profiles and social networks as well as
tracking of user activities, performance, and preferences. In order to reduce
that impact, e.g., due to legal obligations, privacy principles for the purpose
of de-identification\footnote{%
  The often used term of \textit{anonymisation} is avoided, since its usage is
  controversial~\cites{Swe97}[Kap. II.C.2]{Ohm09} and misleading due to the
  associated expectation of absolute privacy.%
}
can be applied such as \textit{generalisation and supression}~\cite{Swe02},
\textit{permutation and aggregation}~\cite{Zha+07},
\textit{perturbation}~\cite{Dwo06}, and \textit{pseudonymisation}~\cite{PH10}.
A pseudonym replaces a QID of a subject in order to prevent or impede its
identification, while at the same time maintains the possibility to re-identify
the subject by means of a certain secret~\cite{PH10}. This provides linkability
of individual \roepl{} as well as allows an investigation of incidents with a
link to individuals, which both are vital requirements for applications such as
intrusion detection and prevention. At the same time, it aims at reducing the
impact, which data collection and processing such as auditing or monitoring has
on privacy by means of data minimization. Therefore, it is a suitable
de-identification technique in the context of event correlation and processing.

However, using pseudonymisation, especially in a distributed setting, as a
technique to provide such a privacy\hyp{}friendly linkability of individual
\roepl{} is not a panacea for absolute data protection and privacy. Several
assumptions have to be made and involved components have to be trusted.
Furthermore, pseudonymisation as a tool to \textit{de-identify} data sets bears
a heavy legacy. Recent history has shown, that even properly de-identified data
can be re-identified with the right background knowledge and so a connection
between data set entries and individuals has been re-established in a long list
of examples~\cites{Swe97, LDM10, Rot10, Mon+13, Sid14, Mon+15, LP16, Dou+16,
CRT17, ED17}.
These recognitions strongly indicate that it is inherently impossible to achieve
full de-identification in general and full privacy via pseudonymisation in
specific. If, however, the term \textit{privacy} is not associated with an
underlying assumption of completeness, but rather is taken in a sense of
reduction and in the best case in a sense of minimisation of QIDs, as it has
been postulated by \textcite{Swe97} and \textcite[Sect.~II.C.2]{Ohm09} already,
pseudonymisation remains a valid tool to enhance the privacy of individuals in
the context of event and data processing. Additionally, no other privacy
preserving technique can be applied to \roepl{} as easily as pseudonymisation,
while preserving the correctness of the original data.
In this regard, it is necessary to move away from a binary distinction between
full privacy on the one hand and no privacy on the other hand, but rather define
privacy as an increasing or decreasing non-formalised continuum.
In view of these facts, our main contributions can be summarised as follows:

\begin{itemize}[nosep]
  \item The explicit formulation of a system and threat model of event and \roe{}
    pseudonymisation and the highlighting of remaining privacy threats,
  \item the proposal of privacy protection goals, which do not aim to achieve
    full privacy, but rather increase privacy by minimisation of personal as
    well as potentially personal data,
  \item the design and development of a pseudonymisation framework including
    technical realisations of all protection goals, and
  \item a discussion about performance implications and an evaluation of the
    framework.
\end{itemize}

The rest of the paper is structured as follows. We provide background on
techniques especially important for our framework in \cref{sec:backgr} as well
as on the system and threat model including possible scenarios of our framwork
and derived requirements in \cref{sec:scn}. The threat model motivates the
proposal of privacy protection goals and a discussion on potential conflicts in
\cref{sec:peepll} including further details about technical realisations of the
privacy protection mechanisms implemented in our event pseudonymisation
framework PEEPLL. We survey related work in \cref{sec:relwork} and conclude in
\cref{sec:concl}.

\section{Background}\label{sec:backgr}

The following sections briefly introduce topics, which are needed as basic
building blocks for the design and implementation of several parts of the
pseudonymisation framework.

\subsection{Pseudonymisation}

The definition of pseudonymisation used throughout this paper is based
on the definition given by GDPR~\cite[]{GDPR16} in Article~4(5):
\enquote{%
  \textit{pseudonymisation} means the processing of personal data in such a
  manner that the personal data can no longer be attributed to a specific data
  subject without the use of additional information [\dots].%
}
As the definition states, some additional information exists, e.g., a mapping
table or a function, to attribute a pseudonym to the corresponding identifier,
which can be used to revert the process of pseudonymisation. We call this
additional information the \pseudomapping{}. Without knowledge about the
mapping, a re\hyp{}identification should be unfeasible without undue effort. The
effectiveness of pseudonymisation is significantly determined by the decision,
which data items constitute QIDs and therefore need to be covered by
pseudonymisation. Note that the process of determining QIDs is higly
application-specific. Note further that the literature often calls the mapping
\enquote{link}, which must not be confused with the term \enquote{Linkability}
explained in \cref{sec:scn:linkability}.

\subsection{Hash Functions and HMACs}
\label{sec:backgr:mac}

A hash function is a mapping \(H\) from an input of arbitrary length to an
output of fixed length \(n \in \mathbb{N}\),
\(
H: M^* \rightarrow M^n,
\)
which is called the hash value or digest of the input. This mapping is
considered unique if the following properties hold:
  \begin{enumerate*}[label=\arabic*)]
    \item \textit{Preimage resistance}: Given the hash value \(y = H(x)\), it is
      infeasible to find the input \(x\) equivalent to invert the hash function
      \(H\).
    \item \textit{Second preimage resistance}: Given an input \(x\), it is infeasible to
      find another input \(x' \neq x\) such that \(H(x') = H(x)\).
    \item \textit{Collision resistance}: It is infeasible to find two arbitrary inputs
      \(x\) and \(x'\) with \(x \neq x'\) such that \(H(x') = H(x)\).
  \end{enumerate*}
Apart from specialised attacks, which target the underlying mathematics and
constructions of hash functions, they suffer from simple brute\hyp{}force
attacks. Especially when the preimage space consist of QIDs, preimage resistance
cannot be provided by hash functions sufficiently~\cite{marx2018-hashing}.

One solution is the enlargement of the preimage space by adding additional
entropy in form of a secret key to the input value. Such a construction is
called a keyed hash function. An HMAC is a practical realisation of a keyed hash
function. As every Message Authentication Code (MAC), it takes a randomly chosen
secret key \(k\) of sufficient length and a message \(m\) of arbitrary length as
input, and outputs \code{Mac(\(k, m\))}, a so called tag of the message, which
is unforgeable as long as the secret key \(k\) is not known~\cite{KL14}.

\subsection{Bloom Filter}
\label{sec:backgr:bloom_filter}

Bloom Filters are used to store and query set member information in a
space\hyp{}efficient way by applying hash functions.
The filter itself consists of a bit array of fixed length \(m\), initially set
to all \(0\)'s. Given \(r\) differing hash functions \(H_i: M^* \rightarrow
\{0,\dots, m-1\} , i \in \{1, \dots, r\}\) mapping inputs to a single bit
position in the filter, a new input can be inserted by computing the \(r\) hash
values for the input and setting the bits at these positions to \(1\).
To test for set membership of an input, the hash values are computed like in the
insert step and the resulting positions are looked up in the filter. If all bits
at the \(r\) positions are set to \(1\), the input is a possible member of the
set.

The space efficiency comes at the cost of possible false positives. If the union
of all member hash values leads to a filter, in which the set bits include all
bits of a non\hyp{}member, the filter would still indicate the set membership
for this non\hyp{}member. On the other hand, a filter, where any bit for the
hash values of an input is set to \(0\), definitely states the
non\hyp{}membership of this input~\cite{bloom1970space}.

\subsection{1-out-of-N Oblivious Transfer}
\label{sec:backgr:oblivious_transfer}

1-out-of-N Oblivious Transfer (OT) refers to a cryptographic primitive, which is
defined as follows: A sender \(\mathcal{S}\) has \(N\) messages \(M_0, \ldots,
M_{N-1}\) and a receiver \(\mathcal{R}\) wants to get the \(i\)-th message
\(M_i\) without \(\mathcal{S}\) learning any information about which message is
of interest for \(\mathcal{R}\). Additionally, \(\mathcal{R}\) shall not
learn any information about any other message \(M_j \neq M_i\) than the
requested one. A very basic 1-out-of-N OT protocol based on the computational
Diffie-Hellman assumption has been proposed by \textcite{NP01} and recently a very
efficient one by \textcite{CO15}:

\begin{description}[nosep,itemindent=\parindent,leftmargin=0pt]
  \item[Preliminaries:] The protocol uses a
    hash function \(H\) and a group \(\mathbb{Z}_p\) of
    prime order \(p\) with \(g\) as a generator of that group.
  \item[Initialisation] (only once, used for all subsequent transfers):
    \begin{enumerate*}[label=\arabic*)]
      \item \(\mathcal{S}\) randomly chooses a secret \(y \in \mathbb{Z}_p\)
        and computes \(s = g^y\) and  \(t = s^y\).
      \item \(\mathcal{S}\) sends \(s\) as its public key to \(\mathcal{R}\).
    \end{enumerate*}
  \item[Input/Output:] \(\mathcal{R}\)'s input is the index \(i \in \{0, \ldots,
    N-1\}\), and \(\mathcal{S}\)'s input is the messages \(M_0, \ldots,
    M_{N-1}\). At the end of the protocol, \(\mathcal{R}\)'s output is
    \(M_{i}\), while \(\mathcal{S}\) learns nothing about \(i\).
  \item[Key Derivation] (for every index of interest for \(\mathcal{R}\), even
    in parallel):
    \begin{enumerate*}[label=\arabic*)]
      \item \(\mathcal{R}\) with input \(i\) randomly chooses a secret \(x
        \in \mathbb{Z}_p\) and computes \(r = s^i \cdot g^x\) as well as \(k_i =
        H(s || r || s^x) = H(s || r || g^{y \cdot x})\).\footnote{Prepending the
          values \(s\) and \(r\) to the hash function as salt approximates a
          random oracle and makes sure that the oracle is local to the protocol
          session. It further helps against malleability attacks \cite{CO15}.}
      \item \(\mathcal{R}\) sends \(r\) to \(\mathcal{S}\).
      \item For all \(j \in \{0, \ldots, N-1\}\) \(\mathcal{S}\) computes \(k_j
          = H(s || r || r^y / t^j) = H(s || r || g^{(y \cdot i + x) \cdot y} /
          g^{y \cdot y \cdot j})\).
    \end{enumerate*}
  \item[Transfer] (for every index of interest for \(\mathcal{R}\), even
    in parallel):
    \begin{enumerate*}[label=\arabic*)]
      \item For all \(j \in \{0, \ldots, N-1\}\) \(\mathcal{S}\) encrypts each
        \(M_j\) by computing \(C_j = Enc(k_j, M_j)\) and then sends these
        encryptions \((C_0, \ldots, C_{N-1})\) to \(\mathcal{R}\).
      \item \(\mathcal{R}\) decrypts \(C_i\) by computing \(M_i = Dec(k_i,
        C_i)\).
    \end{enumerate*}
\end{description}

Research on the topic of OT is manifold and fast-paced. There are many variants
and so called extension, which try to further improve the security or
efficiency. See further~\cite{Ash+17}.

\subsection{Secure Indexes}
\label{sec:backgr:secure_indexes}

Secure Indexes offer the possibility to search for keywords in encrypted
documents by querying specially crafted indexes that maintain the
confidentiality of the indexed keywords~\cite{goh2003-secure-indexes}. Each
Secure Index is based on a Bloom Filter (see \cref{sec:backgr:bloom_filter}),
which encodes the keywords for the corresponding document.
Keywords undergo a two-step encoding before they are inserted in the Bloom
Filter.
\begin{enumerate*}[label=\arabic*)]
  \item A concealment of the keyword by applying a pseudo-random function
\(f\) to both the keyword \(w\) and a secret key \(K = (k_1, \dots, k_r)\).
The output \(x = (x_1 = f(w, k_1), \dots, x_r = f(w, k_r))\) is called trapdoor.
\item A personalisation of each trapdoor with the unique document identifier
\(D_{id}\) by applying the pseudo-random function \(f\) again. The result \(y =
(y_1 = f(x_1, D_{id}), \dots, y_r = f(x_r, D_{id}))\) is called a codeword. Its
elements \(y_i\) are then inserted into the Bloom Filter.
\end{enumerate*}
The second step is to achieve different codewords for identical keywords in
different documents, which avoids a cross-document analysis of common keywords.
The Bloom Filter is stored together with the encrypted document as its Secure
Index. To query if a document contains a given keyword, one calculates the
trapdoor for this keyword, personalises it with the document id, and checks if
the resulting codeword is included in the document's Bloom Filter.
\section{Requirements \& Threat Model}\label{sec:scn}

Before discussing privacy threats and protection goals as well as the design and
implementation of our pseudonymisation framework to mitigate these threats, the
following section introduces principles and requirements for the
pseudonymisation framework and the underlying system model including the most
important basic terms used throughout the rest of the paper. Consider the
following two scenarios, where pseudonymisation of \roepl{} is needed:

\begin{description}[nosep,itemindent=\parindent,leftmargin=0pt]
  \item[Scenario 1\label{scn:1}]%
      An organisation has deployed a distributed security incident detection system
      consisting of several distributed sensors throughout the whole IT as well as
      physical infrastructure in order to monitor user activities from several
      points of view and to detect intrusion, extrusion, and anomalous activities.
      All sensor data is being combined at a central data processing unit, which provides
      abilities to analyse and correlate the data. Most importantly, the data of all
      sensors has to be archived for a certain amount of time to allow a thorough
      investigation in cases of security breaches.

  \item[Scenario 2\label{scn:2}]%
      A consortium of several independent medical institutions wants to collect and
      share medical data on specific rare diseases as well as on their treatment and
      corresponding results in order to be able to improve the quality level of
      treatments especially where one institution does not get enough data by itself
      to achieve statistical relevance.
      The collection and sharing is not limited to a closed set of medical data but
      should be able to incorporate data from patients' follow-up examinations over
      time. Therefore, the data is only useful for research purposes when it is not
      completely de-identified but maintains the ability to update patients' medical
      records from different institutions while the disease status is monitored and
      the treatment is adjusted.
\end{description}

\subsection{Requirements}
\label{sec:scn:requirements}

The two scenarios highlight the following requirements, which need to be
established in order to protect the privacy of individuals in the context of
data collection and sharing.

\subsubsection{Personal Data Minimisation}
\label{sec:scn:data_minimisation}

In both scenarios, the collection, storage and processing of \roepl{} has severe
implications for the privacy of users. To mitigate this effect, the principle of
personal data minimisation should be applied and enforced~\cite{GTD11}.
\enquote{%
  By ensuring that no, or no unnecessary, data is collected, the possible
  privacy impact of a system is limited.%
}~\cite{Hoe14}.
Effective mechanisms for personal data minimisation in these scenarios are
\textit{Select before you Collect}~\cite{Jac05}, which means the limitation of
personal data \enquote{%
  to what is necessary in relation to the purposes for
  which they are processed%
}~\cite[Article~5(1)(c)]{GDPR16} and \textit{Pseudonymisation}. Any other form
of de-identification is not an option for both scenarios, because separate data
items, that are collected, might need to be linked to each other in a way, that
allows the attribution to the same subject, even though it is generally
unimportant, which identity is behind that subject. Additionally, in case of an
investigation in \nameref{scn:1} or a new special treatment possibility for a
patient in \nameref{scn:2}, the subject might need to be re-identifiable.

\subsubsection{Linkability}
\label{sec:scn:linkability}

Monitoring of user, process, and network activities as well as collecting
medical records have not an end in themselves, but are rather embedded into a
broader evaluation process such as intrusion or anomaly detection or statistical
analysis. Linkability~\cite{PH10} provides the context to set individual records
in relation and is the basis for correlation and interpretation.

\subsubsection{Global Pseudonym Consistency}
\label{sec:scn:pseudonym_consistency}

Both scenarios have the requirement, that all \roepl{}, which concern the same
subject, must be pseudonymised such that linkability wrt.\ the subject is
maintained regardless of the data source. Otherwise, analysis and statistics of
the collected and correlated data are distorted. Such a global consistency
requires specific information about already used pseudonyms or a deterministic
algorithm.

\subsubsection{Pseudo- vs.\ Truly-random Pseudonyms}
\label{sec:scn:pseudo}

There are two ways to maintain global pseudonym consistency, both resulting in
different system models.
\begin{enumerate*}[label=\arabic*)]
  \item A pseudonymisation component might derive pseudonyms solely from
    \roepl{} alone and in a deterministic manner. We call this \textit{local
    deterministic} or \textit{pseudo-random pseudonymisation}. This allows the
    setup of a simple pseudonymisation infrastructure, because each
    pseudonymisation component can derive pseudonyms for QIDs independently from
    each other, while at the same time maintain global pseudonym consistency.
    However, \textcite{marx2018-hashing} demonstrated, that preimage attacks via
    brute force to uncover such deterministically derived pseudonyms can be done
    with reasonable effort due to fairly low cardinalities of typical preimage
    domains, such as QIDs (e.g.\ IP addresses, e-mail addresses). Furthermore,
    the need to re\hyp{}identify a subject behind a locally and
    deterministically derived pseudonym, commands the creation, protection, and
    coordination of additional pseudonym disclosure
    information.
  \item Not susceptible to brute force attacks are pseudonyms, which are not
    derived from \roepl{}, but which are chosen truly randomly. To determine
    whether a random pseudonym has already been chosen for a given QID, as it is
    necessary to maintain global pseudonym consistency, a global lookup table
    between QIDs and the corresponding random pseudonyms has to be consulted.
    Re\hyp{}identification of randomly chosen pseudonyms, can be done via this
    lookup table as well.
\end{enumerate*}

\subsubsection{Component Separation}
\label{sec:scn:component_separation}

All information that potentially threatens the privacy of users is collected
decentralised on different independent data sources in both scenarios. For the
deployment of a proper pseudonymisation process, it makes most sense to
pseudonymise all QIDs independently and as close to the data sources as
possible, meaning a one-to-one relation between a data source and a
pseudonymisation component. Furthermore and for the sake of personal data
minimisation a strict separation of all components of the pseudonymisation
process itself should be enforced. On the one hand, data leakage due to a
compromise of a pseudonymisation component does not necessarily break the whole
pseudonymisation process, due to a \textit{separation of knowledge}. On the
other hand, such a separated pseudonymisation allows the exchange of information
between all components taking part in the pseudonymisation process on a so
called \textit{need-to-know} basis. This reduces the exchange and processing of
personal data as well as other potentially jeopardising data, such as pseudonym
frequencies and temporal information about appearances of pseudonyms for
distinct data sources, to an absolute minimum.

\subsubsection{Out of scope}
\label{sec:scn:out_of_scope}

Several other requirements might be important for the process of
pseudonymisation. However, only those mentioned in the previous Sects.\ have
been taken into account for the design of the proposed pseudonymisation
framework. In particular, the actual process for pseudonym disclosure in order
to re\hyp{}identify a subject behind a pseudonym is open for future work.
Furthermore, the problem of authenticity wrt.\ the origin of the \roepl{} will
be considered as out of scope. It is up to the pseudonymisation components to
ensure, that the processed information is authentic. Other requirements might be
the unforgeability or integrity of the \pseudomapping{}, to prevent a malicious
altering of the content of the lookup table, the application of group pseudonyms
and more generally of identical pseudonyms for multiple QIDs,
or the establishment of limited linkability according to a session concept. All
of these have to be established separately, in case they are mandatory
requirements.

\subsection{System Model}
\label{sec:scn:system_model}

\begin{figure}
    \begin{center}
      \includegraphics[width=\linewidth]{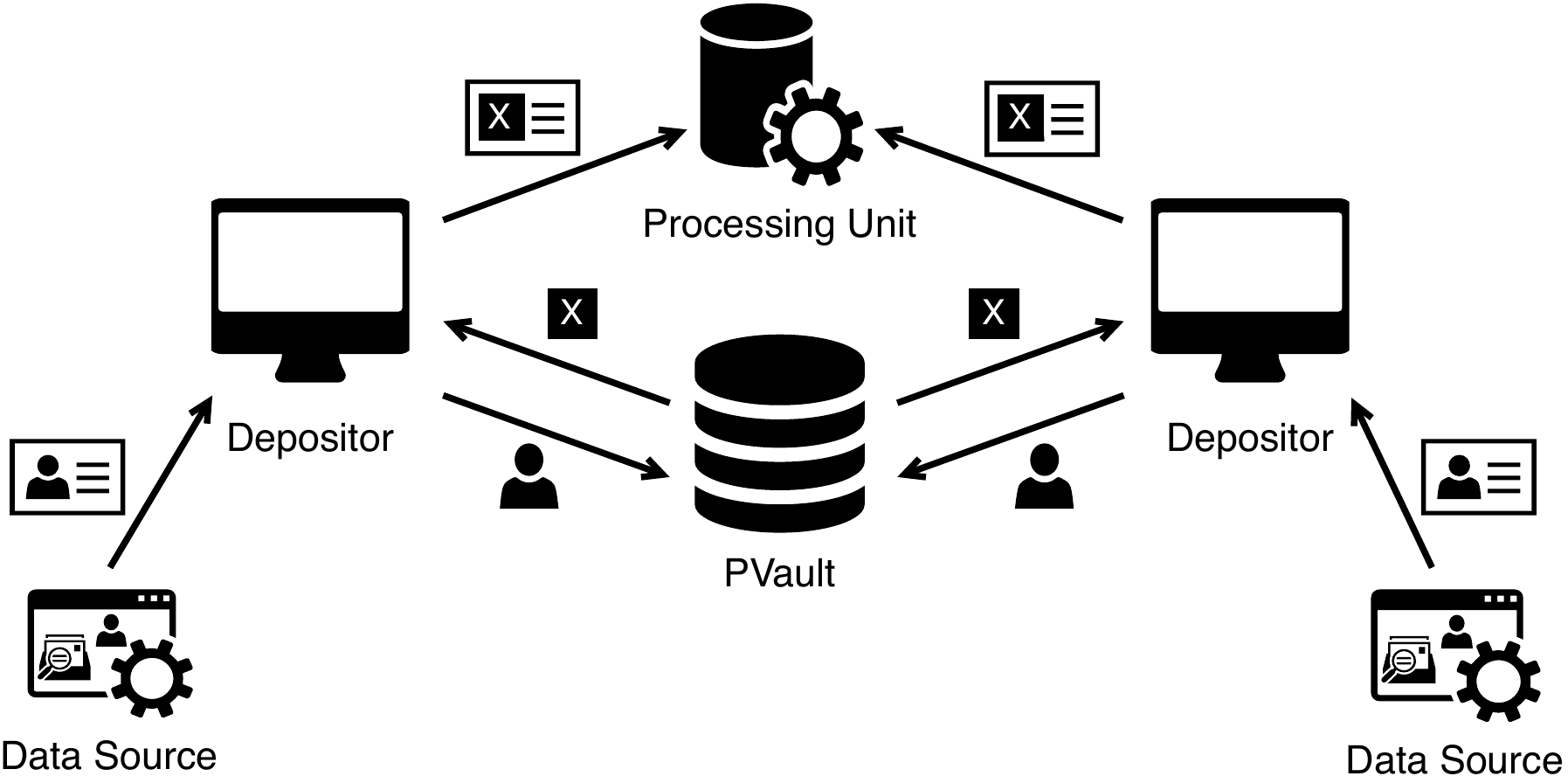}
      \caption{System Model of the Pseudonymisation Framework.}
      \label{fig:system_model}
    \end{center}
    \vspace{-1em}
\end{figure}

The identified and discussed requirements result in the following system model
for our pseudonymisation framework PEEPLL (see \cref{fig:system_model}):
A data source is emitting some representation of an event or some medical
record, which for brevity will be called \roe{} and which possibly contains
personal data of a subject.
A so called \textbf{Depositor} assigned to that data source has the task to
replace a QID of a subject, if present, with a pseudonym. For that, it extracts
QIDs on a single identifier basis one by one, requests a pseudonym for each QID
from a so called \textbf{Pseudonym Vault (PVault)} and upon receiving a response
from the PVault, replaces the identifier with the pseudonym.
The PVault receives a pseudonym request from a Depositor and might be faced with
two situations. First, there already exists a pseudonym for this identifier in
the \pseudomapping{}, which can be sent back to the Depositor. And second, there
does not exist a pseudonym for this identifier in the \pseudomapping{} and so a
new truly-random pseudonym must be created and stored in the \pseudomapping{}
together with the corresponding QID by the PVault.
The pseudonymised \roe{} is then sent from the Depositor to the data collection
and correlation unit.

Remarks: The integrity and confidentiality of all communication is protected.
The pseudonymisation process is
transparent to the data sources and the processing unit. The PVault is needed in
order to provide truly-random pseudonyms while at the same time maintain global
pseudonym consistency. Communication for the
pseudonymisation process only is needed between the Depositors and the PVault,
not between the Depositors themselves.\footnote{%
  Even though the Depositors provide identical functionality and interact
  identically with the PVault, they can not be seen as one logical component,
  because each Depositor should be limited to the scope of its data source.%
}

\subsection{Threat Model}
\label{sec:scn:threat_model}

PEEPLL does not aim at providing provable privacy against strong external or
internal attackers. In fact, there are several threats, which directly undermine
the privacy protection mechanisms of PEEPLL, mainly because of their fundamental
nature:

\begin{itemize}
  \item A malicious data source can undermine the pseudonymisation process for
    all locally processed \roepl{}.

  \item A malicious Depositor can
    \begin{enumerate*}[label=\arabic*),labelsep=0pt,itemindent=!,leftmargin=0pt]
      \item \label{itm:1}create his own local \pseudomapping{} containing all
        locally processed QIDs and pseudonyms, as well as
      \item \label{itm:2}ignore pseudonym responses or even
        skip requesting pseudonyms from the PVault and use his own (random or
        not) pseudonyms.
    \end{enumerate*}

  \item A malicious PVault can manipulate the pseudonym lookup of already
    processed QIDs as well as manipulate the generation function and ignore the
    requirement of random pseudonyms.
\end{itemize}

Apart from organisational rules and regulations, there are no protection
mechanisms, which could be deployed by PEEPLL in order to prevent these threats.
At most, a weak mitigation against these fundamental privacy threats can be
achieved by a strict separation of all components and an enforcement to prevent
a collution, since it is limiting their scope to the locally accessible data.
Nonetheless, from a threat model point of view, these components need to be
considered as trusted and a collaboration must be prohibited. Furthermore,
recent publications on re-identification of de-identified data strongly
indicate, that it is inherently impossible to achieve full privacy via
pseudonymisation (see \cref{sec:intro}).

However, PEEPLL moves away from a binary distinction between full privacy on the
one hand and no privacy on the other hand, but rather sees privacy as well as an
opposing threat to privacy as an increasing or decreasing non-formalised
continuum. With this association of the term \textit{privacy}, the minimisation
of existing identifying and quasi-identifying features in a system exacerbates
the privacy threats mentioned above and increases the privacy of individuals.
Conversely, the existence of such QIDs including meta data of a pseudonymisation
process itself constitutes a threat to privacy, since it aids to re-identify
individuals.
Thus, the main focus of PEEPLL is the strict minimisation of QIDs, including
meta data of the pseudonymisation process itself, such as re-usage patterns of
pseudonyms. It tries to maintain the information that really is needed by the
data processing unit in order to provide its functionality as well as the
information that really is needed for the process of pseudonymisation. The
remaining sources of potentially sensitive information shall be eliminated as
accurately as possible. In particular, PEEPLL facilitates the minimisation of
the following data or the prevention of its utilisation:

\begin{itemize}
  \item Meta information about the usage count of existing pseudonyms can be
    inferred by Depositors, which will be prevented by PEEPLL with specifically
    designed responses of the PVault so that Depositors do not learn such
    information (see \cref{sec:peepll:reuse_ind}).
  \item The PVault has a global view on all cleartext QIDs prevalent in the
    whole system. This threat will be addressed in PEEPLL by protecting the
    confidentiality of all QIDs with respect to the PVault (see
    \cref{sec:peepll:deposit_conf}).
  \item The PVault can infer usage patterns from pseudonym requests as a form of
    meta information, which, over time, might provide the ability to infer
    additional sensitive information. PEEPLL will mitigate this threat by
    limiting the linkability of data records (see
    \cref{sec:peepll:limited_linkability}) as well as by preventing the PVault
    from learning any information about which entry of the pseudonym lookup
    table matches a queried deposit (see \cref{sec:peepll:match_pseudo_unobs})
  \item The collection of pseudonymised \roepl{} in general allows the creation
    of individual profiles over pseudonyms, which increase in accuracy over
    time, since all \roepl{} related to the same QID are linkable via the
    corresponding pseudonym. Those profiles, even without being directly
    linkable to individuals, might on the one hand leak sensitive information
    and on the other hand illegally be de-pseudonymised, which both have an
    increasing success probability over time. PEEPLL will mitigate this threat
    by limiting the linkability of several related \roepl{} (see
    \cref{sec:peepll:limited_linkability}).
\end{itemize}

\section{PEEPLL}\label{sec:peepll}%
\label{sec:design}%

Pseudonymisation alone does not protect against tracking,
profiling, and re-identification via background knowledge attacks
\cite{BZH06,Toc14}.\footnote{%
  See also \cref{sec:intro} and \cref{sec:scn:threat_model}.%
}
We will consider the following protection goals with respect to the design and
implementation of the proposed pseudonymisation framework. Each protection goal
aims at reducing the impact on the privacy of individuals. The overall challenge
is to preserve the linkability of certain \roepl{} to some specifiable extent.

\subsection{Re-use Indistinguishability}
\label{sec:peepll:reuse_ind}

\begin{definition}
  The information about whether or how often a pseudonym has been used before by
  any Depositor is only known to the PVault. Especially, a Depositor should not
  be able to distinguish whether or not a pseudonym has been used before by any
  other Depositor.
\end{definition}

A pre\hyp{}requisite of this protection goal is the separation of pseudonym creation
from querying existing or new pseudonyms, which confirms the
importance of the distributed environment already established by the system
model.
From a technical point of view to protect the \nameref{sec:peepll:reuse_ind},
the PVault is responding to a pseudonym request of a Depositor in such a way,
that both cases, the generation of a new pseudonym and the finding of a matching
entry in the \pseudomapping{}, look identical and thus are indistinguishable to
the Depositor. This approach, however, is not applicable to every configuration
of PEEPLL. See \cref{sec:peepll:match_pseudo_unobs:new_data_item} for more
details.

\subsection{Deposit Confidentiality}
\label{sec:peepll:deposit_conf}

\begin{definition}\label{dfn:deposit_conf}
  The QID, which is to be replaced with a pseudonym by a
  Depositor, is only known to that Depositor itself. Neither the PVault nor any
  other Depositor shall learn any information about the underlying QID
  from a pseudonym request or a deposit, except another Depositor is
  processing the same QID as well.
\end{definition}

PEEPLL utilises HMACs by equipping all Depositors with a shared secret \(k\) not
known to the PVault. Given \(k\) and a \(QID\), which is to be pseudonymised, a
Depositor converts the plain \(QID\) into a lookup token \(T_{QID} =\)
\code{Mac(\(k, QID\))} calculated by the tag\hyp{}generation function of the
HMAC\@. This lookup token will later be used to recognize a possibly
corresponding existing pseudonym \(P_{QID}\) without relying on information
about the \(QID\) itself -- therefore establishing
\textit{\nameref{sec:peepll:deposit_conf}}. Furthermore, because all Depositors
use the same shared secret \(k\), the lookup token of \(QID\) stays consistent
across the whole system, which preserves the requirement of
\textit{\nameref{sec:scn:pseudonym_consistency}}. The Depositor sends a pseudonym request
containing \(T_{QID}\) to the PVault owning the global \pseudomapping{} \(PM\),
which is a set of pairs of existing lookup tokens and corresponding pseudonyms
\(\{(T_{QID_1}, P_{QID_1}), \dots, (T_{QID_n}, P_{QID_n})\}\) for already
processed QIDs. The PVault searches \(PM\) for a
matching entry \((T_{QID}, P_{QID})\) of the lookup token and the corresponding
pseudonym connected to \(QID\). In case such a matching pair in \(PM\) does not
exist yet, the PVault has to generate a new pseudonym \(P_{QID}\), store it
together with the lookup token in \(PM\) and respond the Depositor with
\(P_{QID}\). The Depositor replaces the \(QID\) with the received pseudonym
\(P_{QID}\) and carries on. Compare \cref{fig:design-mac}.

\begin{figure}
    \begin{center}
        \begin{tikzpicture}
        \input{images_arxiv/message_flow_mac}
        \end{tikzpicture}
        \caption{Interaction between a Depositor and the PVault in the
          HMAC\hyp{}based approach.}
        \label{fig:design-mac}
    \end{center}
    \vspace{-1em}
\end{figure}

The secret key, which is shared among all Depositors but not with the PVault,
adds entropy to the hashing process, which is necessary since the PVault could
otherwise easily brute\hyp{}force the actual QID (see
\cref{sec:backgr:mac}). In PEEPLL it is generated and distributed to all
Depositors once at the very beginning of the deployment.
Such a brute\hyp{}force attack is still possible, in cases where Depositors get
hold of foreign deposits not related to their currently processed QID\@. This
can happen when protection mechanisms for
\textit{\nameref{sec:peepll:match_pseudo_unobs}} (see
\cref{sec:peepll:match_pseudo_unobs}) are deployed as well. Since all Depositors
know the secret key, the additional entropy is truncated. We refer to this
problem as \textit{Weak \nameref{sec:peepll:deposit_conf}} and discuss a
solution in \cref{sec:peepll:deposit_conf_and_match_pseudo_unobs:ot}.

\subsection{Matching Pseudonym Un\-ob\-ser\-va\-bi\-li\-ty}%
\label{sec:peepll:match_pseudo_unobs}

\begin{definition}
  Which pseudonym from the \pseudomapping{} actually matches a specific data
  item requested by a Depositor is only known to the Depositor itself. In other
  words, the PVault does not learn any information about which entry of the
  \pseudomapping{} matches a queried deposit.
\end{definition}

It does not matter if the QIDs, which need to be
pseudonymised, can be processed as plaintexts or if they need to be concealed in
order to protect the \nameref{sec:peepll:deposit_conf}. Because of this, in the
following, they are denoted as data item \(E\).
In its most simple form, this protection goal can be achieved by sending the
whole \pseudomapping{} to each Depositor who requests a pseudonym. In this way,
the PVault does not learn which entry is of real interest. However, besides the
need of a great amount of bandwidth, this approach raises two problems.

\subsubsection{Privacy Issue}
\label{sec:peepll:match_pseudo_unobs:privacy_issue}

All existing data items as well as their corresponding pseudonyms will become
known to the requesting Depositor, which contradicts the principle of
\textit{\nameref{sec:scn:data_minimisation}} and the limitation of the scope of
one Depositor. PEEPLL balances this conflict and additionally saves bandwidth by
limiting the number of returned \pseudomapping{} entries, while at the same time
assures, that this number is truly greater than one.\footnote{%
  It must be highlighted, that this approach achieves
  \textit{\nameref{sec:peepll:match_pseudo_unobs}}, but does not achieve
  \textit{\nameref{sec:peepll:deposit_conf}}, even when the confidentiality of
  data items is protected by the approach explained in
  \cref{sec:peepll:deposit_conf}, since a Depositor getting hold of foreign
  deposits is able to brute\hyp{}force the HMACs. For a solution of this
  \textit{Weak \nameref{sec:peepll:deposit_conf}} see
  \cref{sec:peepll:deposit_conf_and_match_pseudo_unobs:ot}%
}

A Depositor converts the data item \(E\) into a lookup token \(T_E\) and sends a
pseudonym request containing \(T_E\) to the PVault owning the global
\pseudomapping{} \(PM\). The lookup token creation is returning a filter or
mask, which can be applied to the \pseudomapping{} \(PM\) by the PVault and
which matches both the entry of real interest as well as other irrelevant
entries.
In particular, the lookup token \(T_E\) created by a Depositor for a specific
data item \(E\) consists of a Bloom Filter (see~\cref{sec:backgr:bloom_filter}),
which not only contains the data item \(E\), but also a blinding of \(b\)
randomly chosen bits. The blinding accomplishes an artificial false positive
rate, which influences the probability, that more than one entry matches a given
lookup token, while searching the \pseudomapping{} \(PM\). The false positive
rate can be controlled via the number of blinding bits \(b\) (see
\cref{sec:eval_false_positive_rates}).
The PVault applies the lookup token to \(PM\) and returns a set of all matching
pairs of pseudonyms and the corresponding data items. The Depositor itself
searches the set for a matching pair \((E_j, P_{E_j})\), where \(E_j = E\),
replaces the data item \(E\) with the pseudonym \(P_{E_j}\) and carries on.

\subsubsection{New Data Item Issue}
\label{sec:peepll:match_pseudo_unobs:new_data_item}

The second problem relates to the case of a new data item of real interest,
which does not exist in the \pseudomapping{} yet. In this case, the returned set
only contains irrelevant entries, which will and must be sorted out by the
Depositor. Eventually, the Depositor has to request the creation of a new
pseudonym from the PVault, thereby thwarting
\textit{\nameref{sec:peepll:reuse_ind}} and also
\textit{\nameref{sec:peepll:match_pseudo_unobs}}, since the creation request
unambiguously references the data item of real interest.
This can only partially be fixed by forcing the creation of dummy pseudonyms.
After each pseudonym request and its corresponding response, the Depositor must
send a pseudonym creation request, which either contains the data item of real
interest, when only irrelevant \pseudomapping{} entries have been sent back from
the PVault, or which contains a dummy data item otherwise.
\textit{\nameref{sec:peepll:reuse_ind}} will still be violated, since the Depositor
can distinguish meaningful from irrelevant responses. This problem remains open
for future work.

\subsection{Combined \nameref*{sec:peepll:deposit_conf} and
\nameref*{sec:peepll:match_pseudo_unobs}}
\label{sec:peepll:deposit_conf_and_match_pseudo_unobs}

In order to protect both the \textit{\nameref{sec:peepll:deposit_conf}} and
\textit{\nameref{sec:peepll:match_pseudo_unobs}} simultaneously, the concept of
1-out-of-N OT (see \cref{sec:backgr:oblivious_transfer}) in conjunction with
HMACs as utilised in \cref{sec:peepll:deposit_conf} seems to be a natural
solution.
However, OT is not easily applicable to our context for two
reasons. First, it assumes, that the database is of fixed length and its entries
can be accessed by specific indices known to the Depositor. But the
\pseudomapping{} of the PVault is constantly changing and its entries as well as
their indices are not known to the Depositor. Second, OT involves heavy
computation or heavy communication or both. At the end of the OT protocol, the
sender delivers the whole database (specifically encrypted) to the receiver,
which is not suitable to our setting. For this reason, the mechanisms explained
in \cref{sec:peepll:deposit_conf} and \cref{sec:peepll:match_pseudo_unobs} are
combined in PEEPLL as an alternative to OT. Additionally, certain optimisations
are applied, which provide a more tailored solution and which closely relate to
the concept of Secure Indexes (see \cref{sec:backgr:secure_indexes}). The
concept of OT still does provide a viable solution to the problem of
\textit{Weak \nameref{sec:peepll:deposit_conf}}, so its application in PEEPLL is
discussed in \cref{sec:peepll:deposit_conf_and_match_pseudo_unobs:ot}.

\subsubsection{Secure Indexes paired with HMACs}
\label{sec:peepll:deposit_conf_and_match_pseudo_unobs:si_mac}

\begin{figure}
    \begin{center}
        \begin{tikzpicture}
        % \draw[help lines] (0,0) grid (5,-6);

% Depositor
\node[scale=4] (Depositor) at (0,0) {\faDesktop};
\node[below=-1em of Depositor] (dep-desc) {Depositor};
\draw (dep-desc.south) -- ++(0,-5);

% Idvault
\node[scale=4] (IDVault) at (4.5,0) {\faDatabase};
\node[below=-1em of IDVault] (idv-desc) {PVault};
\draw (idv-desc.south) -- ++(0,-5);

% messages
\draw[->] (0.1,-2.35) -- ++(0.3,0);
\draw (0.5,-2.1) rectangle ++(3.5,-0.5) node[midway] {\(T_{QID} = f(k, QID)\)};

\draw[<-] (4.1,-3.45) -- ++(0.3,0);
\draw (0.5,-3.2) rectangle ++(3.5,-0.5) node[midway] {\(\{(MAC_{QID_1}, P_{QID_1}), \dots \}\)};

\draw[->, dotted] (0.1,-4.75) -- ++(0.3,0);
\draw[dashed] (0.5,-4.5) rectangle ++(3.5,-0.5) node[midway] {\((MAC_{QID},
BF_{QID})\)};

\draw[<-, dotted] (4.1,-5.85) -- ++(0.3,0);
\draw[dashed] (0.5,-5.6) rectangle ++(3.5,-0.5) node[midway] {\(P_{QID}\)};

% algorithms
\node[fill=white] at (0, -1.75) {\code{LookupToken}};
\node[fill=white] at (4.5, -2.9) {\code{SearchMapping}};
\node[fill=white] at (0, -4.2) {\code{[CreateIndex]}};
\node[fill=white] at (4.5, -5.3) {\code{[UpdateMapping]}};
        \end{tikzpicture}
        \caption{Interaction between a Depositor and the PVault in
          the Secure\hyp{}Index\hyp{}-based approach.}
        \label{fig:design-sec-idx}
    \end{center}
    \vspace{-1em}
\end{figure}

Given the shared secret \(k\), a Depositor creates a Bloom Filter as lookup
token \(T_{QID} = BF(k, QID)\) for the data item \(QID\). The Bloom Filter is
constructed by deriving a set of \(r\) secret keys \((k_1, \cdots, k_r)\) from
\(k\), randomly picking a subset of \(r/2\) keys and repeatedly applying a
pseudo\hyp{}random function to the \(QID\) for each \(k_i\) in the subset,
determining which bits in the Bloom Filter are set (see
\cref{sec:backgr:bloom_filter}). Using only half of the keys during lookup
introduces an indeterminism to mitigate query profiling as discussed by
\textcite{goh2003-secure-indexes} (compare
\cref{sec:eval_false_positive_rates}).
The lookup token then is sent to the PVault, who owns the \pseudomapping{}
\(PM\), which is a set of triples each consisting of a Bloom Filter, a HMAC, and
a corresponding pseudonym for already processed QIDs. The PVault  returns a set
\( \{(HMAC_{QID_j}, P_{QID_j}) \,\,\, | \,\,\, (BF_{QID_j}, HMAC_{QID_j},
P_{QID_j}) \in PM \wedge T_{QID} \subset BF_{QID_j}\} \) of all pairs of HMACs
and pseudonyms whose respective Bloom Filter is a superset of \(T_{QID}\). The
expected number of elements of the set is influenced by the false positive rate
introduced by the blinding injected into the Bloom Filters of the
\pseudomapping{} (see \cref{sec:eval_false_positive_rates}).

In order for a Depositor to recognize the proper pseudonym in the received
result set, the locally computed \(HMAC_{QID} =\) \code{Mac(\(k, QID\))} is
compared to the received \(HMACs\). If no match is found, the Depositor takes
the \(HMAC_{QID}\) and creates a Bloom Filter similar to the one created as
lookup token with the exception that all \(r\) keys are used instead of a
subset, and adds a blinding of \(b\) bits to the Bloom Filter by setting \(b\)
randomly chosen bits to \(1\). The blinding accomplishes an artificial false
positive rate, which influences the probability, that more than one triple
matches a given lookup token while searching the \pseudomapping{} \(PM\). The
false positive rate can be controlled via the number \(b\) of blinded bits. The
result is a pair \((HMAC_{QID}, BF_{QID})\) of the \(HMAC\) and Bloom Filter,
which correspond to the \(QID\). This is sent to the PVault, who updates the
\(PM\) with the resulting pair \(BF_{QID}, HMAC_{QID}\) and a newly generated
pseudonym \(P_{QID}\), which is finally returned to the Depositor. This
interaction is shown in \cref{fig:design-sec-idx}.

Note: The presented approach protects \textit{\nameref{sec:peepll:reuse_ind}}
and \textit{\nameref{sec:peepll:match_pseudo_unobs}} only in those cases, where
the Depositor finds a matching HMAC in the result set from the PVault. For
further details, see \cref{sec:peepll:match_pseudo_unobs:new_data_item}.

\subsubsection{False Positive Rates}
\label{sec:eval_false_positive_rates}

Given an observed event rate \(r\), the defined retention period \(p\) and an
aspired false positive rate \(fp\), the number of unique words can be
approximated as \(n = r*p*c\), where \(c\) is a constant to account for the
expected number of identifiers per event.
According to Goh~\cite{goh2003-secure-indexes}, Bloom Filters should be
parameterized with a number of hash functions \(k = -\log_2 fp\) and a Bloom
Filter size of \(m = \nicefrac{n*k}{\ln 2}\).

Since we employ Goh's extension proposal for a heuristic, that obfuscates query
duplicates by sending only a partial trapdoor, the effective false positive rate
is elevated.
Instead of using the full trapdoor of length \(k\), a random sample of size
\(k'=k/2\) is drawn from the full trapdoor.
Consequently, the resulting false positive rate is given as \(fp' = 2^{-k'} =
2^{-k/2} = \sqrt{2^{-k}} = \sqrt{fp}\).
To compensate for the effect of partial trapdoor building on the effective false
positive rate, the number of hash functions is chosen as \(k^* = -2 \log_2 fp\),
where \(fp\) is the desired false positive rate.

\begin{figure}
    \begin{center}
        \begin{tikzpicture}
        \begin{axis}[
    width=0.3\textwidth,
    xlabel=false positive rate \(fp'\),
    ylabel=number of matches]
    \addplot[smooth,mark=*,blue] plot coordinates {
       (0.03162277660168379, 5.02)
       (0.07071067811865475, 10.32)
       (0.1, 12.28)
       (0.15811388300841897, 16.24)
       (0.22360679774997896, 27.98)
       (0.27386127875258304, 25.86)
       (0.31622776601683794, 23.52)
       (0.3535533905932738, 47.16)
       (0.3872983346207417, 46.58)
       (0.4183300132670378, 46.44)
       (0.4472135954999579, 48.82)
    };
    %\addlegendentry{Simple}
\end{axis}
        \end{tikzpicture}
        \caption{Average number of matching Bloom Filters dependent on the false
          positive rate measured by adding deposits to a prefilled PVault (\(100\)
          records).}
        \label{fig:eval_fp_rate}
    \end{center}
    \vspace{-1em}
\end{figure}
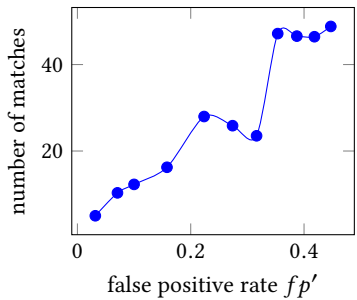

We have measured this relation in the deployed implementation displayed in
\cref{fig:eval_fp_rate}.
It shows the average number of deposits with matching Bloom Filters dependent on
the false positive rate \(fp'\) for a fixed number of pre-added deposits.
Obviously our approach can only give probabilistic guarantees for the number of
false positives and as a consequence of this for the protection goal of
\nameref{sec:peepll:match_pseudo_unobs}.

\subsubsection{Solving \textit{Weak \nameref{sec:peepll:deposit_conf}}}
\label{sec:peepll:deposit_conf_and_match_pseudo_unobs:ot}

The presented approach still raises a special challenge. The entries of the
\pseudomapping{} consist of pairs of confidential QIDs and corresponding
pseudonyms, with the mandatory requirement, that the technical realisation to
protect the \textit{\nameref{sec:peepll:deposit_conf}} is of deterministic nature. This
determinism assures, that a QID always results in the same concealed data item
(the HMAC of a QID), which is sent to the PVault as a pseudonym request and
stored in the \pseudomapping{} for future reference. This determinism
additionally assures, that a Depositor, who receives a set of deposits as a
response to its pseudonym request, which potentially contains deposits from
earlier pseudonym requests as well as from other Depositors, and which should
remain confidential even to this Depositor (see \cref{dfn:deposit_conf}), is
able to recognise the one deposit, which matches the actual QID of real
interest.
However, this determinism makes all deposits, or more precisely their HMACs,
that are returned to the Depositor, susceptible to brute\hyp{}force attacks by
that Depositor, because the additional entropy added to the HMACs is known to
the Depositor and so is useless as protection mechanism against an enumeration
of all possible QIDs\@. As a result, the deposits, which are to be sent to the
Depositor as a response to a pseudonym request, have to be processed in a way,
that conceals all deposits except the one of real interest for the Depositor,
before leaving the PVault.

A perfectly valid solution is the application of 1-out-of-N OT as explained in
\cref{sec:backgr:oblivious_transfer} with some adjustments to overcome its two
major obstacles for our setting.
\begin{enumerate*}[label=\arabic*)]
  \item The first obstacle concerns the missing fixed indices, which would need
    to be used to pinpoint a specific entry of the \pseudomapping{} on the side
    of a Depositor playing the receiver of the OT protocol. A resolution is the
    utilisation of the HMACs themselves, which are part of the lookup tokens, as
    keys or indices of a hash map storing the corresponding entries of the
    \pseudomapping{}. As this again would deliver the HMACs of foreign deposits
    to a Depositor, who can easily brute\hyp{}force them, the HMACs as indices
    of the hash map are hashed a second time together with the OT specific entry
    key formerly denoted as \(k_j\) for all \(j \in \{0, \ldots, N-1\}\) with
    \(N\) being the number of all entries: \(OT\mbox{-}INDEX_{QID_j} =\)
    \code{Mac(\(k_j, HMAC_{QID_j}\))}. The Depositor who requests the pseudonym
    for a specific QID can calculate the OT-key \(k_{QID}\), which is used to
    decrypt the requested deposit out of the received set of encrypted deposits.
    This OT-key also allows the Depositor to compute the correct
    \(OT\mbox{-}INDEX_{QID}\) and so identify the requested deposit in the received set.
  \item The second obstacle concerns the computational as well as the
    communication overhead introduced by the OT protocol. This obstacle can at
    least be mitigated by not using the whole \pseudomapping{} as input for the
    sender (PVault), but limiting the number of inputs to the ones requested by
    the Depositor via the Bloom Filter. This denotes a trade-off between
    achievable security of OT and the performance of the
    pseudonymisation framework.
\end{enumerate*}

\subsection{Limited Linkability}
\label{sec:peepll:limited_linkability}

\begin{definition}
  The linkability of \roepl{} concerning the same QID should
  only be maintained for a specified and limited period.
\end{definition}

Limiting the linkability constitutes a trade-off between the requirements of
\textit{\nameref{sec:scn:data_minimisation}} and
\textit{\nameref{sec:scn:linkability}}. This trade-off has to be optimised by
adjusting variables to control the limitation, which is a highly
application\hyp{}specific process.
Technical mechanisms to achieve
\textit{\nameref{sec:peepll:limited_linkability}} are realised in PEEPLL by
limiting the time period in which reuses are possible as well as by limiting the
re-uses.

\subsubsection{Temporal Limitation by Global Epochs}%
\label{sec:design:temporal_limitation}

Temporally limiting the linkability of pseudonyms is realised in PEEPLL by
introducing epochs, at whose beginning all pseudonyms are changed. The PVault is
enforcing the limitation by simply deleting the existing \pseudomapping{}. On
the Depositor's side, temporal limitation is achieved by combining the \(QID\)
with an epoch specific tag \(t_i\) in a way, that the PVault cannot link lookups
for \(QID \circ t_i\) and \(QID \circ t_j\), where \(i \neq j\), to the same
\(QID\). For that, PEEPLL utilises HMACs in the same way as they are use in
order to achieve \textit{\nameref{sec:peepll:deposit_conf}}. The epoch specific
tag \(t_i\) for epoch \(i\) is deterministically derived from the secret key
\(k\) that is shared among all Depositors as a master secret. Given this epoch
tag and a \(QID\), which is to be pseudonymised, the combination \(QID \circ
t_i\) is derived by converting the plain data item into a lookup token
\(T_{QID_{t_i}} =\) \code{Mac(\(t_i, QID\))}.
A major advantage of this approach is the possibility, that it can be enforced
by both the Depositor and the PVault in such a way that no \reusage{} is
possible beyond limitation if at least one party complies. We will refer to this
two-sided enforcement as \textit{anytrust}.

\subsubsection{Budget Limitation}

By limiting the linkability of \roepl{} by a budget, it is not possible to
re-use a pre\hyp{}existing pseudonym mapping after the budget sum of prior
re\hyp{}uses has exceeded the maximum privacy budget. The individual budgets
are either simply \(1\) or a weight that is sensitive to the context and the
impact of a pseudonym re\hyp{}use on the privacy of a subject.
In its most simple form, the privacy budget accumulates the number of re-uses of
a pseudonym. Such an approach can be achieved by introducing usage counters for
each pseudonym on the PVault. If the counter for a pseudonym exceeds the upper
bound, the existing entry in the \pseudomapping{} is deleted or hidden, which
triggers the creation of a new pseudonym for this QID on its
next request.
%
\iffalse%
\paragraph*{Achieving Anytrust Without Loosing \nameref*{sec:peepll:reuse_ind}}

A possible solution would be to employ secure multi\hyp{}party computation
protocols.
\fi
%
In order to prevent the revelation of the actually matching pseudonym by the
budget accounting to the PVault when enforcing
\textit{\nameref{sec:peepll:match_pseudo_unobs}}, the costs for the current
request is added to all pseudonyms that match the lookup including the
irrelevant matches. As a consequence, the budget counter associated to a
pseudonym on the PVault would only be a fuzzy account and an upper bound of its
actual budget.

\subsection{Achievable Protection Goals}

In the preceding section, we have introduced technical means for several of our
protection goals, namely \textit{\nameref{sec:peepll:reuse_ind}},
\textit{\nameref{sec:peepll:deposit_conf}},
\textit{\nameref{sec:peepll:match_pseudo_unobs}},
\textit{\nameref{sec:peepll:deposit_conf_and_match_pseudo_unobs}} as well as
\textit{\nameref{sec:peepll:limited_linkability}}. However, some of these
technical means lead to violations of other protection goals. Especially, our
approaches for achieving \textit{\nameref{sec:peepll:match_pseudo_unobs}} are
built in such a way that the protection goal of
\textit{\nameref{sec:peepll:reuse_ind}} is not achieved. Finding solutions which
accomplish these combinations of protection goals should be seen as an open
challenge.

\section{Related Work}\label{sec:relwork}

The concept of pseudonymisation and \textit{Limited Linkability} of pseudonymous
data has attracted some interest in concrete research fields, especially
vehicular ad hoc networks (VANETs) and mobile communication.

\textcite{petit2015} survey different strategies for pseudonym creation in
VANETs and highlight changing pseudonyms as a vital requirement in the lifecycle
of a pseudonym. The risk of enabling an attacker to create mobility patterns of
drivers has to be balanced with the linkability required for operational tasks
like collision detection. However, in opposition to VANETs, where one vehicle
uses a pseudonym, our scenario has the requirement of \textit{Global Pseudonym
Consistency} for data receiving different pseudonyms coming from different data
sources, so that the given strategies are not applicable.

\textcite{arapinis2014} examine the handling of pseudonym updates in mobile
communication networks, namely the strategy of reallocating temporary
identifiers (TMSI) in the 3GPP standard. They find that current implementations
of the standard miss reasonable reallocation strategies and therefor enable user
tracking for long time periods over different areas and independent of the
amount of user activity. We try to prevent similar tracking approaches in our
scenario by limiting the linkability as given in
\cref{sec:peepll:limited_linkability}.

Recently \textcite{florian2015sybil} proposed a pseudonymisation and pseudonym
change approach utilizing the blockchain technique to achieve a complete
decentralisation and resistance against sybil attacks. While this approach is of
interest for pseudonymous authentication challenges, e.g. in the vehicular
communication, it is not suitable for our scenario of multiple data sources
requesting pseudonyms for different data items.

In intrusion detection research pseudonymisation has been widely examined to
balance automatic anomaly detection requirements with privacy requirements.

In 1997, \textcite{sobirey_pseudonymous_1997} discussed the impact, which the
collection and analysis of audit events might have on users' privacy and
presented pseudonymisation as a viable way to protect the privacy interests of
employees and users of computer systems in general against the upcoming trend of
automatic intrusion detection and operating system auditing. They describe the
distributed intrusion detection system \textit{AID} which uses deterministic
encryption to provide consistent pseudonyms for different agents in their
system. However, they do not discuss the consequences of \textit{Global
Pseudonym Consistency} in a distributed system as well as issues with
\textit{Limited Linkability}.

\textcite{buschkes_privacy_1999}~describe the conflicting interests of an IDS
operator and the monitored users in detail. They demand the concepts of data
avoidance and data minimization. They propose a pseudonym\hyp{}based solution,
which requires a central trusted third party knowing the user identities for
pseudonym generation. To minimize the impact of user profiling they introduce
the concept of group reference pseudonyms referencing user groups instead of
single users. Our approach uses \textit{Component Separation} to prevent the
central component from learning user identities.

\textcite{biskup_threshold-based_2000,biskup_transaction-based_2000} substitute
identifying features in audit event messages with transaction pseudonyms, which
are derived via Shamir's secret sharing from a longer\hyp{}living pseudonym. If
the number of audit events concerning the same identity, i.e.\ the number of
issued secret shares, exceeds a defined threshold, an auditor can recover the
pseudonym from the observed shares. The correlation of pseudonyms generated by
different pseudonymisation components, as we try to achieve with
\framework{PEEPLL}, is stated as an open issue.

\section{Conclusion and Outlook}\label{sec:concl}

This paper works out sustained privacy threats to the pseudonymisation of
\roepl{} due to the existence of identifying and quasi-identifying data as well
as meta data of the pseudonymisation process itself, proposes four privacy
protection goals, and presents privacy enhancements for the application of
pseudonymisation of \roepl{} in form of a framework. We first provided the
motivation based on two real-world scenarios, where identifying and
quasi-identifying material needs to be processed and the privacy of the affected
subjects can be restored by utilising pseudonymisation. Based on these
scenarios, we identified three important requirements for our event
pseudonymisation, namely \textit{\nameref{sec:scn:pseudonym_consistency}},
\textit{\nameref{sec:scn:component_separation}}, as well as
\textit{\nameref{sec:scn:linkability}}, and formalised our setting with a
concrete system and threat model in \cref{sec:scn}. This includes an explanation
of operational limits to the achievable level of privacy by event
pseudonymisation, which lead to the formulation of the four privacy protection
goals \textit{\nameref{sec:peepll:reuse_ind}},
\textit{\nameref{sec:peepll:deposit_conf}},
\textit{\nameref{sec:peepll:match_pseudo_unobs}}, and
\textit{\nameref{sec:peepll:limited_linkability}} in \cref{sec:peepll}. An
important observation is the sometimes contrary nature of these protection goals
and the resulting conflicts and obstacles, which arise when two or more of these
protection goals shall be achieved. For each protection goal, the technical
realisation in PEEPLL as well as potential performance consequences were
explained as well. Assuming all components of our framework act compliantly to
their protocol (honest-but-curious), the framework provides the following
properties:

\begin{itemize}[nosep]
  \item Pseudonymisation with \textit{\nameref{sec:scn:pseudonym_consistency}},
  \item Enforcing \textit{\nameref{sec:peepll:limited_linkability}} (temporal and budget)
  \item Protection of \textit{\nameref{sec:peepll:deposit_conf}},
  \item Protection of \textit{\nameref{sec:peepll:match_pseudo_unobs}}
  \item Protection of \textit{\nameref{sec:peepll:reuse_ind}}, but only in
    combination with \textit{\nameref{sec:peepll:deposit_conf}}, not with
    \textit{\nameref{sec:peepll:match_pseudo_unobs}}.
\end{itemize}

\noindent
Practical scenarios for the application of our framework are not limited to the
ones mentioned in \cref{sec:scn}, but those were the focus of our development.
PEEPLL will be applied to those scenarios in the context of two of our research
projects, where the requirement of data analysis and monitoring meets strict
privacy regulations. The first context is the privacy respecting detection and
prevention of insider attacks, which embraces extensive monitoring of employee
activities and so attacks from one of the pseudonymisation components themself
(PVault, Depositors) have to be taken into account. The second context is the
collection and processing of medical patient data, which deals with highly
privacy relevant records on the one hand, and the need to analyse detailed
information from patients' disease processes on the other hand.

Future work will be conducted on the integration of a pseudonym
re-identification (pseudonym disclosure) process and potential side effects on
the achievement of the privacy protection goals. Further aspects comprise the
\textit{Weak Deposit Confidentiality}, the enforcement of \textit{anytrust} in
the context of \textit{Budget Limitation}, and the simultaneous protection of
\textit{\nameref{sec:peepll:reuse_ind}} and
\textit{\nameref{sec:peepll:match_pseudo_unobs}}.

\iffalse%
\begin{acks}
  Part of this work has been developed in the project DREI, which is partly
  funded by the \grantsponsor{BMBF}{German Federal Ministry of Education and
  Research}{https://www.bmbf.de/en/} under the reference
  number~\grantnum{BMBF}{16KIS0545}.
\end{acks}
\fi

\begin{raggedright}     %
  \renewcommand*{\bibfont}{\footnotesize} %
  \printbibliography
\end{raggedright}

\end{document}